\documentclass[sigconf]{acmart}
\AtBeginDocument{%
  \providecommand\BibTeX{{%
    \normalfont B\kern-0.5em{\scshape i\kern-0.25em b}\kern-0.8em\TeX}}}

\setcopyright{acmcopyright}
\copyrightyear{2022}
\acmYear{2022}
\acmDOI{XXXXXXX.XXXXXXX}

\usepackage[noend]{algpseudocode}
\usepackage[ruled]{algorithm2e}
\usepackage{subcaption}

\begin{document}

\title{Transferable Fairness for Cold-Start Recommendation}

\author{Yunqi Li}
 \affiliation{%
  \institution{Rutgers University, New Brunswick, NJ, USA}
  \country{}
    }
 \email{yunqi.li@rutgers.edu}
 
 \author{Dingxian Wang}
 \authornote{This work completed when Dingxian Wang worked at eBay}
 \affiliation{%
  \institution{eBay Global Growth, Seattle, WA, USA}
    \country{}
    }
 \email{dingxianwang@gmail.com}
 
\author{Hanxiong Chen}
\authornote{This work completed when Hanxiong Chen worked as a PhD at Rutgers University}
 \affiliation{%
  \institution{Rutgers University, New Brunswick, NJ, USA}
    \country{}
    }
 \email{hanxiong.chen@rutgers.edu}

\author{Yongfeng Zhang}
 \affiliation{%
  \institution{Rutgers University, New Brunswick, NJ, USA}
    \country{}
    }
 \email{yongfeng.zhang@rutgers.edu}




\begin{abstract}

With the increasing use and impact of recommender systems in our daily lives, how to achieve fairness in recommendation has become an important problem. Previous works on fairness-aware recommendation mainly focus on a predefined set of (usually warm-start) users. However, recommender systems often face more challenging fairness issues for new users or cold-start users due to their insufficient amount of interactions. Therefore, it is essential to study whether the trained model still performs fairly for a new set of cold-start users. This paper considers the scenario where the recommender system meets new users who only have limited or even no interaction with the platform, and aims at providing high-quality and fair recommendations to such users effectively. The sufficient interaction data from warm users is treated as the source user domain, while the data from new users is treated as the target user domain, and we consider to transfer the counterfactual fairness from the source users to the target users. To this end, we introduce a framework to achieve transferable counterfactual fairness in recommendation. The proposed method is able to transfer the knowledge of a fair model learned from the source users to the target users with the hope of improving the recommendation performance and keeping the fairness property on the target users. Experiments on two real-world datasets with representative recommendation algorithms show that our method not only promotes fairness for the target users, but also outperforms comparative models in terms of recommendation performance.

\end{abstract}



\maketitle

\section{Introduction}
Modern recommender systems built on collaborative filtering techniques have achieved great success in many applications. However, they still suffer from a variety of important challenges such as the cold-start problem, i.e., how to provide good recommendation services to new users \cite{ricci2011introduction}. Existing methods to solve such problems are mostly based on introducing auxiliary information to help to learn the representations of cold users or items, such as items' affinity information \cite{hong2019crowdstart}, users' personal features \cite{lika2014facing}, and graph structures \cite{lu2020meta}. Some other works consider solving the cold-start issue through data augmentation \cite{hwang2016told}. However, existing research on cold-start recommendation only focus on the recommendation accuracy of the target cold-start users, but neglect the fairness of the recommendations provided to the target cold-start users.
Yet, fairness is an important perspective to consider in recommender systems so as to achieve responsible recommendation \cite{zehlike2022fairness, deldjoo2022fairness, ekstrand2022fairness, wang2022survey, yao2017beyond, leonhardt2018user, beutel2019fairness, li2021user, ge2021towards, li2022fairness}.


Previous works of fairness-aware recommendation mainly focus on designing a fair model for a pre-defined set of users with sufficient training data, while overlooking the cold-start new user issue that is commonly encountered in practice. Therefore, it is essential to study whether recommender systems still perform fairly for cold-start users, and how we can transfer fairness from warm users to cold users. In this paper, we target at the cold-start scenario where there are new users coming into the system with limited or even no interaction. We treat the new users together with their interactions as the target user domain, while the warm-start users with sufficient interaction data are treated as the source user domain. Our goal is to learn the shared property between the source and target users and transfer the knowledge of a fair model trained from the source users to the target users to maintain the user fairness as well as improving their recommendation performance.

Some recent works have noticed the necessity to transfer fairness among different domains for classification tasks \cite{yoon2020joint, schumann2019transfer, madras2018learning}. The existing works mainly focus on when and how we can transfer fairness defined on association notions, such as demographic parity \cite{Zafar2015LearningFC}, equalized odds \cite{hardt2016equality}, and equal opportunity \cite{hardt2016equality}, with the aim of mitigating the discrepancy of statistical metrics between individuals or sub-populations. However, it has been shown by recent research that fairness cannot be merely assessed based on association notions \cite{khademi2019fairness,kusner2017counterfactual,zhang2018equality,zhang2018fairness}, since they can not reason about the causal relationships between the protected features (such as gender) and the predicted outcomes (such as hiring results). Therefore, causality-based fairness notions have been proposed and have become more and more important in modelling fairness in machine learning \cite{khademi2019fairness,kusner2017counterfactual,zhang2018equality,zhang2018fairness}. The advantage of causality-based fairness notions is that they are usually defined based on the causal model to leverage prior knowledge about the world structure, and thus can help to understand the causal relationship between variables in the system as well as how they change. Therefore, in this paper, we consider to achieve transferable counterfactual fairness between source users and target users.

Technically, we introduce a framework for transferring knowledge together with counterfactual fairness from source users to target users. The framework can be divided into two steps. For the first step, we introduce a framework to achieve counterfactually fair recommendation for the source user. To meet the requirement of counterfactual fairness, we show that we need to make sure the predicted recommendation lists are independent from the users' sensitive features. To this end, we integrate adversary learning and the training of recommendation models to generate feature-independent user representations, while keeping those representations informative enough for the recommendation task. For the second step, we transfer the knowledge and counterfactual fairness from source users to target users. Traditional transfer learning approaches usually transfer knowledge by learning a mapping function for user representations \cite{man2017cross}. However, such methods have limitations in practical use since they have to perform transfer learning based on the overlapped users between source and target. In this paper, we aim at a more practical but more challenging case where there are no overlapped users between the source and target, and we
take advantage of generative adversary networks \cite{goodfellow2014generative} to perform knowledge transfer. Our method provides the flexibility to consider both the unsupervised case where target users have no interaction, and the supervised case where target users have several limited interactions. We also provide a theorem to show how we can transfer counterfactual fairness based on our two-step framework. Experiments on two real-world datasets with some representative recommendation algorithms show that our method is able to transfer both fairness and recommendation performance from the source users to the target users.

The key contributions of this paper are as follows:

\begin{itemize}
     
    \item We consider to achieve transferable fairness for users in recommendation task. To better assess fairness, we consider to transfer counterfactual fairness from source users to target users.
    
    \item We introduce a two-step framework for transferring knowledge and counterfactual fairness from a learned fair model of the source users to the target users based on adversarial learning.

    \item We conduct experiments on two real-world datasets with both shallow and deep base models to show the effectiveness of our framework on transferring both recommendation performance and fairness across users.
\end{itemize}

In the following, we review related work in Section \ref{sec:related}. Some preliminaries and notations are provided in Section \ref{sec:notations}. We introduce the details of our framework in Section \ref{sec:framework} and Section \ref{sec:framework2}. Experimental settings and results are provided in Section \ref{sec:experiments}. Finally, we conclude this work in Section \ref{sec:conclusions}.

\section{Related Work}
\label{sec:related}

\subsection{Fairness in Recommendation}
Fairness in recommendation has become an important topic. Since recommender systems involve multiple stakeholders, the fairness concerns in the recommendation tasks can be put forward from very different perspectives. Some works concern the unfairness issue for items and focus on the popularity bias/unfairness problem in recommendation, i.e., the popular items will get more exposure opportunity than those less popular ones. Such problem usually can be solved by increasing the number of unpopular items or otherwise the overall catalog coverage in the final recommendation list \cite{adomavicius2011improving,kamishima2014correcting,abdollahpouri2017controlling,abdollahpouri2019managing}. For example, \citeauthor{beutel2019fairness} \cite{beutel2019fairness} propose a pairwise comparative ranking framework to achieve item fairness in recommendation by adding a regularizer into the objective function. There are also some works considering user fairness in recommendation. Examples include Lin et al. \cite{xiao2017fairness}, which consider user fairness under group recommendation scenario and propose an optimization framework based on Pareto Efficiency; \citeauthor{leonhardt2018user} \cite{leonhardt2018user}, which achieve fairness for users by post-processing algorithms and improve the diversity in recommendation; and \citeauthor{li2021user} \cite{li2021user}, which require the recommender system to treat the groups of active and inactive users similarly. Some other works see fairness in recommendation tasks from multi-sided view, such as \citeauthor{burke2017multisided}
\cite{burke2017multisided} and  \citeauthor{abdollahpouri2019multi} \cite{abdollahpouri2019multi}, which consider unfairness issue under multi-stakeholder scenario and introduce several corresponding group fairness properties;
\citeauthor{mehrotra2018towards}
\cite{mehrotra2018towards}, which jointly optimize fairness and performance in two-sided marketplace platforms; and \citeauthor{patro2020fairrec} \cite{patro2020fairrec}, which explore individual fairness in two-sided platforms from the view of long-term sustainability. 
Recently, few works pay attention to exploring unfairness issues in recommendations with cold-start users. \citeauthor{wu2022big} \cite{wu2022big} study the problem of whether big recommendation models perform fairly to cold-start users and propose a BigFair method based on self-distillation. \citeauthor{wei2022comprehensive}\cite{wei2022comprehensive} study ensuring user fairness of meta-learned recommendation models under cold-start cases. The aim and scope of our work differ from theirs as we consider improving the fairness of cold-start users in general collaborative filtering recommendation which is not limit to big or meta-learned models. To the best of our knowledge, our work is the first to address transferable fairness from warm-start source users to cold-start target users in recommender systems.

\subsection{Adversary Learning in Recommendation}
Transfer learning has been shown effective to alleviate the data sparsity problem in recommender systems \cite{li2009transfer, pan2013transfer, zhao2013active}. Previous works usually consider the problem as a domain adaptation problem with the assumption that the sparse data from target users are somehow similar to the sufficiently labeled data from source users. Generative Adversarial Network (GANs) \cite{goodfellow2014generative}, which is known as a successful method in adversarial learning, have recently been integrated to solve the domain transfer problem in recommendation. For example, IRGAN \cite{wang2017irgan} uses GAN in the context of item-based recommendation. RecSys-DAN \cite{wang2019recsys} explores GAN in cross-domain recommender systems by adopting multi-level generators and discriminators for user features, item features and their interactions. 
CnGAN \cite{perera2019cngan} uses GAN to generate cross-network user preferences for non-overlapped users, and regards the learning process as a mapping from the target to the preference manifold of the source network. \citeauthor{yan2020cross} \cite{yan2020cross} generate adversarial examples dynamically to improve the generalization ability of the cross-domain recommendation model. \citeauthor{li2020atlrec}~\cite{li2020atlrec} consider the overlapped users between target and source users and transfer shareable features across them. \citeauthor{li2021recguru} \cite{li2021recguru} propose a framework based on GAN to generate generalized user representations across domains in sequential recommendation when there is minimum or no overlapped user across domains. Our work differs from those works as we not only aim to transfer useful information from source users to target users to improve the recommendation performance, but also transfer fairness between source and target users.

\subsection{Transferable Fairness}
Some recent work have begun to study the transfer learning of fairness metrics among different domains. \citeauthor{yoon2020joint}~\cite{yoon2020joint}
propose a method for developing a fair classification model under unsupervised case by transferring knowledge from the domain with enough information. 
\citeauthor{schumann2019transfer}~\cite{schumann2019transfer} provide a theoretical understanding of how and when the group fairness in one domain transfers to another in classification tasks. \citeauthor{coston2019fair}~\cite{coston2019fair} focus on situations where protected
attributes are not available in either the source or target domain. \citeauthor{madras2018learning}~\cite{madras2018learning} explore adversarial learning to learn representations which are used by third parties with unknown objectives, and consider group fairness such as demographic parity, equalized odds, and equal opportunity to different adversarial objectives. \citeauthor{lan2017discriminatory}~\cite{lan2017discriminatory} observe that standard transfer learning can improve the prediction accuracy of target tasks but will not transfer fairness well. Our work differs from the previous works in that we consider to transfer fairness in recommendation task with cold-start users, and we focus on transferring counterfactual fairness rather than association-based fairness metrics between source and target to better assess unfairness issues.

\section{Preliminaries}
\label{sec:notations}

In this section, we introduce the preliminaries and notations about recommendation task and counterfactual fairness. In this paper, capital letters such as $X$ denote variables, lowercase letters such as $x$ denote specific values of the variables. Bold capital letters such as $\textbf{X}$ denote a set of variables, while bold lowercase letters such as $\textbf{x}$ denote a set of values. 

\subsection{Recommendation Task}

In general recommendation tasks, we usually have $n$ users and $m$ items denoted as a user set $\mathbb{U}=\left\{u_{1}, u_{2}, \cdots, u_{n}\right\}$ and an item set $\mathbb{V}=\left\{v_{1}, v_{2}, \cdots, v_{m}\right\}$, as well as the user-item interaction history denoted as a 0-1 matrix $Y =\left[y_{i j}\right]_{n \times m}$, where each entry $y_{ij}=1$ if user $u_i$ has interacted with item $v_j$, otherwise $y_{ij}=0$. Modern collaborative filtering-based recommender systems usually take user-item interaction history as input to learn user embedding $\textbf{r}_u$ and item embedding $\textbf{r}_v$ for user $u$ and item $v$, and calculate the preference score $S_{uv}$ based on the embeddings so as to generate a top-$N$ recommendation list to user $u$ with the $N$ items having the highest preference scores. 
To solve the cold-start recommendation issue through transfer learning, we consider two different types of users called the source user $\mathcal{D}_S$ and target user $\mathcal{D}_T$, where the source user $\mathcal{D}_S$ contains the users who have sufficient data of interactions for training a high-quality recommendation model, while the target user $\mathcal{D}_T$ contains users with limited or even no interaction. Therefore, we have $\mathcal{D}_S \cap \mathcal{D}_T = \varnothing$, and we assume the items are shared between source and target users. What's more, we use $\textbf{A}$ to represent the sensitive attributes of users such as age and gender, and use $\textbf{X}$ to denote all the insensitive attributes of users, i.e., the attributes which are not causally dependent on $\textbf{A}$.

\subsection{Counterfactual Fairness}
Counterfactual is known as a "what if" statement where the "if" portion is unreal or unrealized \cite{pearl2016causal}. For example, job seekers usually care about the fairness issue on gender, that is, "what the hiring decision will be if I were a male/female?" We use counterfactual to compare two outcomes under the exact same condition, differing only in the "if" portion. Counterfactual fairness is an individual-level fairness notion defined based on counterfactual \cite{kusner2017counterfactual}, which requires that the final prediction results for every possible individual are the same in the counterfactual world as in the real world. In this paper, we consider the counterfactual world as the one where users' sensitive features have been changed, while all the other features that are not casually dependent on the sensitive features are kept unchanged.  For example, suppose we need a recommender system that does not discriminate against gender when it decides the admission of candidates to a university. Counterfactual fairness requires that the admission result to a male/female candidate will not be changed if his/her gender were reversed while all other attributes that are not dependent on gender such as GPA remain the same. Since students of different genders will have different preferences for majors and future jobs, it is easy to see that the requirement of counterfactual fairness can be more reasonable than those association-based fairness notions which usually forcefully require the same admission rate for all genders. To be clearer, the definition of counterfactually fair recommendation is as follows.

\begin{definition}[Counterfactually fair recommendation]\label{counterfactual fairness}
A recommender model is counterfactually fair if for any possible user $u$ with insensitive features $\textbf{X}=\textbf{x}$ and sensitive features $\textbf{A}=\textbf{a}$, for all the Top-N recommendation lists $L$ for user $u$, and for any value $\textbf{a}^{\prime}$ attainable by $\textbf{A}$, we have:
$$
P\left(L_{\textbf{a}} \mid \textbf{X}=\textbf{x}, \textbf{A}=\textbf{a}\right)=P\left(L_{\textbf{a}^{\prime}} \mid \textbf{X}=\textbf{x}
, \textbf{A}=\textbf{a}\right)
$$

\end{definition}

The concept of counterfactual fairness is proposed by \citeauthor{kusner2017counterfactual} \cite{kusner2017counterfactual}, and here we generalize the definition to recommendation scenario. This definition requires that the distribution of the generated recommendation results $L$ for a given user $u$ should be the same if we only change $\textbf{A}$ from $\textbf{a}$ to $\textbf{a}^{\prime}$, while holding the remaining features $\textbf{X}$ unchanged. Here the notation $P\left(L_{\textbf{a}^{\prime}} \mid \textbf{X}=\textbf{x}
, \textbf{A}=\textbf{a}\right)$ involves two worlds: the observed world where $\textbf{X}=\textbf{x}$ and $\textbf{A}=\textbf{a}$ and the counterfactual world where $\textbf{X}=\textbf{x}$ and $\textbf{A}=\textbf{a}^\prime$. The expression represents the distribution of recommendation results $L$ had $\textbf{A}$ been $\textbf{a}^{\prime}$ given that we observed $\textbf{X}=\textbf{x}$ and $\textbf{A}=\textbf{a}$. We can see that the counterfactual fairness can be achieved if the sensitive features \textbf{A} are independent of the recommendation results $L$ for every user.

\section{Fairness-aware Recommendation}
\label{sec:framework}
In Section \ref{sec:framework} and Section \ref{sec:framework2}, we introduce our framework for achieving transferable counterfactual fairness in recommendation (TFR). The framework can be split into two sub-frameworks. Firstly, we introduce the framework for generating counterfactually fair recommendations for source users in Section \ref{sec:framework}. In Section \ref{sec:framework2}, we show how we can transfer fairness as well as recommendation performanceto the target users. The whole framework is designed as the combination of two sub-frameworks to bring the two following notable advantages. First, in real-world scenarios, recommender systems are usually designed to be very complicated, and the amount of data from the warm users for training the model is pretty huge. Therefore, it is time-consuming to retrain the whole model frequently to generate recommendations for the relatively small number of new users. To solve the problem, our method transfers the knowledge learned from the original huge model and data without modifying it. Second, we make no assumption of the underlying recommendation model so as to offer the model-agnostic flexibility. Our framework can be built upon any model as long as it learns user representations for making recommendations, which is the most popular working mechanism of modern recommender systems.

\subsection{Problem Formulation}
We first analyze how we can achieve counterfactual fairness in recommendation. To satisfy Definition~\ref{counterfactual fairness}, we need to make sure that the generated recommendation results are independent of users' sensitive features.

Modern recommender systems usually learn representations of users and items from user-item interaction histories. For a given user $u$, the system takes the embeddings of the user $\textbf{r}_u$ and corresponding candidate items $\textbf{r}_v$ to generate the recommendation list for user $u$.
Although the sensitive information of users may not be used as input to train the model, it still can be encoded into the user embeddings through the learning process due to already biased training data \cite{li2021towards}. 
For example, female users may have taken more low-income jobs than male users due to unfair social discrimination and pay gap, and such biased user-job pairs are logged into the training data.
As a result, during the collaborative learning process, even though the model does not directly use gender as a feature for model training and recommendation, such unfair ``female user -- low-income job'' relationship could be implicitly captured and encoded into the learned user embeddings, which leads to unfair recommendation results. Therefore, to meet the requirement of counterfactual fairness, we need to guarantee the independence between user embedding $\textbf{r}_u$ and their sensitive features $\textbf{A}_u$, i.e., we need to ensure $\textbf{r}_u \perp \textbf{A}_{u}$, for all $ u \in \mathcal{U}$. 

\subsection{Method}

To satisfy counterfactual fairness, we take advantage of adversary learning to learn feature-independent user embeddings. Many explorations have been made on the use of adversary learning to mitigate discrimination by removing the information of sensitive features from the representations \cite{elazar2018adversarial, wang2019balanced, bose2019compositional,arduini2020adversarial,du2020fairness}. The main idea is to train a filter module together with an adversarial discriminator module simultaneously. The filter module aims to filter out the information of sensitive features from representations, while the goal of the discriminator module is to predict the sensitive features from the representations. 
For our first step, we follow the adversary learning setup \cite{goodfellow2014generative,bose2019compositional,arduini2020adversarial, li2021towards} and develop an adversary network that consists a filter module and a discriminator module.  

\subsubsection{\textbf{Filter Module}}
We first introduce the filter module with a filter function. The filter function is denoted as $f_\textbf{A}: \mathbb{R}^{d} \mapsto \mathbb{R}^{d}$, and the aim of the filter function is to filter out the information of certain sensitive features from user embeddings $\textbf{r}_u$. 
We denote the filtered embedding as $\textbf{r}_u^* = f_\textbf{A}(\textbf{r}_u)$, which is independent from certain sensitive features $\textbf{A}$ while maintaining other insensitive information of the user. Here for brevity, we consider the case that $\textbf{A}$ contains a single sensitive feature. When there are multiple sensitive features to consider, our method can be easily generalized by building a set of filter functions.

\subsubsection{\textbf{Discriminator Module}}
To learn a strong filter function, we train a discriminator module with an adversarial classifier inspired by the idea of adversary learning. The goal of the discriminator is to distinguish the sensitive features from user embeddings. In specific, for the sensitive feature $A$, we train a classifier $D_{A}: \mathbb{R}^{d} \mapsto$ [0,1], which attempts to predict $A$ from the user embeddings. The training process tries to jointly optimize both goals of filter and discriminator until the discriminator can not tell the label of sensitive feature from user embeddings.

\subsubsection{\textbf{Adversary Training}}

To achieve counterfactual fairness in recommendation, we optimize the recommendation loss together with the adversary loss. $\mathcal{L}_A$ denotes the loss of discriminator $D_{A}$, and we use a cross-entropy loss here for implementation. $\mathcal{L}_{Rec}$ denotes the loss of the recommendation task, which varies with different base recommendation models and can take different forms such as the pair-wise ranking loss \cite{rendle2012bpr} and mean square error loss \cite{koren2009matrix}. Therefore, the adversary learning loss of our first step is as the following:
\begin{equation}\label{eq}
\mathcal{L}_1=\sum_{u, v \in \mathcal{D}_S}\left(\mathcal{L}_{\text {Rec}}(f_A(\textbf{r}_u),\textbf{r}_v, y_{uv})
-\lambda_A \cdot \mathcal{L}_A \left(f_A(\textbf{r}_u), A\right)\right)
\end{equation}
Here the adversarial coefficient $\lambda_A$ controls the trade-off between recommendation performance and fairness. A larger $\lambda_A$ usually means a stricter requirement of fairness, and we may need to scarifice more recommendation performance to meet the fairness demands. In an extreme case when $\lambda_A \rightarrow \infty$, the filter function will learn a trivial solution that always output a constant. We will get a very fair recommendation result but totally lose the accuracy of the recommendation performance.

\section{Transferable Fairness}
\label{sec:framework2}
In this section, we talk about the details of how to transfer the knowledge and fairness from source users to target users. From the first step we introduced in the previous section, we could get a fairness-aware recommendation model with informative and fair user embeddings for source users, as well as the well-trained item embeddings shared across users. In our second step, we propose to learn high-quality embeddings for target users through transfer learning, while fixing the item embeddings and other network parameters learned from the source users, so as to generate recommendations for the target users.

\begin{figure*}[t!]
    \centering
    \includegraphics[scale=0.55]{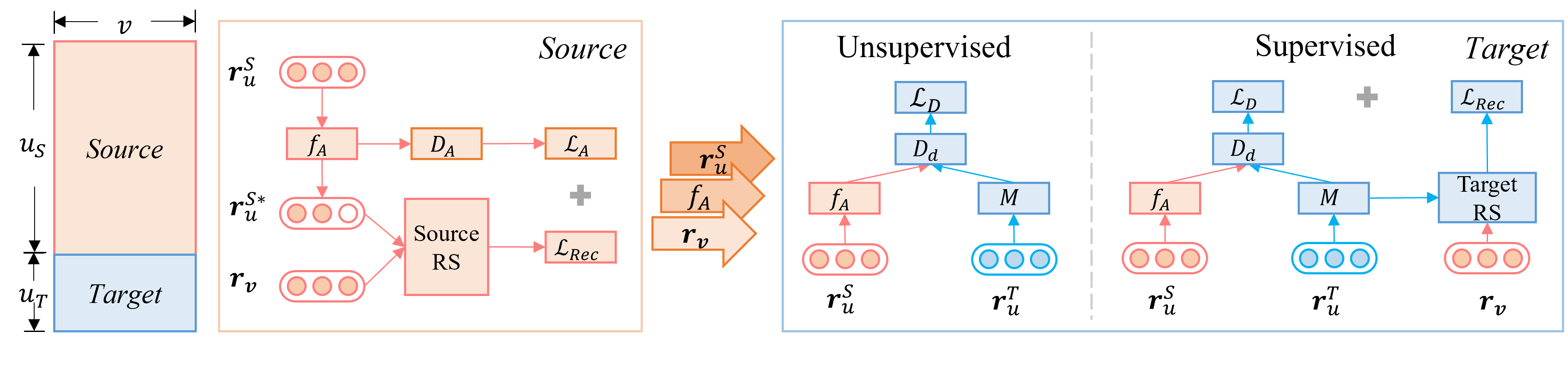}
    \caption{Illustration of our framework. The leftmost figure shows our transfer setting is for non-overlapped users but shared item set. We use $red$ color to represent source user components and use $blue$ color to represent target user components. We first train a fairness-aware recommender system for the source users, and then transfer the knowledge and fairness learned from source users to the target users through adversarial learning.}
    \label{fig:overview}
    \vspace{-10pt}
\end{figure*}

 
Our approach is motivated by the domain adaptation theory that a good representation to transfer is the one for which an algorithm cannot identify the original domain of the input data \cite{ben2006analysis}. 
In the first step, we have already got informative and fair user embeddings for source users. Now we aim to learn high-quality embeddings for target users by mitigating the distance between the distributions of embeddings across source and target users through adversary learning. 


\subsubsection{\textbf{Unsupervised Case}} We first consider the unsupervised case where the target users have no interactions available to use. To mitigate the distribution distance of user embeddings between source and target, we train a mapping function $M: \mathbb{R}^{k} \mapsto \mathbb{R}^{d} $ to map the user embeddings of target users into the same distribution of user embeddings of source users. Here the input of the mapping function can be anything helpful to train recommender models including user ID, user features, as well as randomly initialized or pre-trained embeddings. In our implementation, we use randomly initialized embeddings as input to test the model performance under extreme cases where no interaction and metadata is available, so as to show the effectiveness of transfer learning. To learn an effective mapping function, we also take the idea of adversary learning to train a discriminator $D_d: \mathbb{R}^{d} \mapsto$ [0,1]. The mapping function $M$ aims at mitigating the distance between distributions across the source and target so that we can get informative representations for target users, while the adversarial discriminator aims to predict the original domain of the input user embeddings. The training process tries to jointly optimize both goals until the discriminator can not tell the domains of user embeddings. The training loss is:
\begin{equation}
   \mathcal{L}_{\text {D}} =\min_M\max_{D_d}\sum_{u \in \mathcal{D}_S \cup \mathcal{D}_T}\log \left( D_d\left(f_A\left(r_u^S\right)\right)\right)+\log \left(1-D_d\left(M\left(r_u^T\right)\right)\right)
\end{equation}

The intuition of the loss function is to play a min-max game between the discriminator $D_d$ and the mapping function $M$. When the mapping function is fixed, the loss is trying to learn a discriminator that can best distinguish the target user embedding distribution from the source distribution; while when fixing the discriminator, the loss is trying to learn a mapping function that minimizes the distance between the source and target distributions, so that the target user embedding distribution will finally converge to that of source users.


\subsubsection{\textbf{Supervised Case}} We also consider the supervised case where the target users have limited interactions. In some real-world scenarios, there will be a few available interactions of the target users, our framework provides the flexibility to add this information into the training process to generate more accurate recommendations. Similar to the unsupervised case, we also train a mapping function 
$M$ and a discriminator $D_d$ to mitigate the distance between distributions across domains. Besides, we train a recommendation loss using the interaction data together with the adversary learning to improve the recommendation accuracy. The training loss is:
\begin{equation}
   \mathcal{L}_2 = \sum_{u, v \in \mathcal{D}_T}\mathcal{L}_{\text {Rec}}(u, v, y_{uv}) + \lambda_D \cdot \mathcal{L}_{\text {D}}
\end{equation}

Figure~\ref{fig:overview} shows the whole architecture of our framework. The two-steps algorithm is shown in Algorithm~\ref{alg}, which is in Appendix~\ref{ap:alg}.

We provide the following theorem to show the theoretical guarantee that our two-steps framework can achieve transferable counterfactual fairness across domains in recommendation task. The proof can be seen in Appendix~\ref{ap:proof}.

\begin{theorem}
\label{theorem1}
If (1) the mapping function and discriminator are implemented with sufficient capacity, and (2) at each step of Algorithm~\ref{alg} (shwon in Appendix~\ref{ap:alg}), the discriminator is allowed to reach its optimum given the mapping function, and (3) the mapping function is optimized according to the loss function with discriminator fixed, then after the two steps of Algorithm~\ref{alg}, we have for any user $u \in \mathcal{D}_T$, $\textbf{r}_u \perp \textbf{A}_{u}$ as $\lambda\rightarrow \infty$. 
\end{theorem}

\section{Experiment}
\label{sec:experiments}
In this section, we conduct experiments to show the effectiveness of our method on both transferring counterfactual fairness and recommendation performance. Before we analyze the experiment results, we first introduce the data, baseline models as well as evaluation metrics. The detailed experimental settings can be seen in Appendix \ref{ap:exp}.

\subsection{Dataset}
\subsubsection{\normalfont \textbf{Description}}
We conduct experiments on two publicly available real-world recommendation datasets which contain the information of users' sensitive features to analyze fairness. The statistics of the two datasets are provided in Table~\ref{tb:data}.

\subsubsection*{\normalfont \textbf{MovieLens-1M\footnote{https://grouplens.org/datasets/movielens/1m/}}} This is a commonly used benchmark dataset which contains user-item interactions and user profile information for movie recommendation. We treat \textit{gender}, which is a binary value feature, as the sensitive feature. We convert the explicit ratings into implicit feedback by replacing the observed 1 to 5 ratings into "1"s and unseen interactions as "0"s.

\subsubsection*{\normalfont \textbf{Last.FM}\footnote{http://www.cp.jku.at/datasets/LFM-2b/}} This dataset contains musician listening information from Last.fm online music system. The original dataset is collected by~\citeauthor{melchiorre2021investigating}~\cite{melchiorre2021investigating}, and contains 2 billion interactions for doing research about fairness on gender in recommendation. We follow~\cite{melchiorre2021investigating} by randomly down sampling 10\% users with their interactions to build a smaller dataset. For each user, we also treat \textit{gender} as the sensitive feature, where \textit{gender} is a binary value feature.

Please refer to Appendix~\ref{ap:data} for details about how we split data into source and target, and further divide each dataset for training, validation and testing.



\begin{table}[t!]
  \caption{Statistics of the datasets}
  \begin{tabular}{lcccc}
    \toprule
    Dataset& \#Interactions & \#Users & \#Items & Sparsity  \\
    \midrule
     MovieLens-1M & 1,000,209& 6,040& 3,952&95.81\%\\
     Last.FM &693,660 &5,434 &8,844 &98.56\%\\
  \bottomrule
 \label{tb:data}
\end{tabular}
\vspace{-20pt}
\end{table}

\vspace{-10pt}
\subsection{Baselines}
It is worth noting that our proposed method is a framework that aims to improve the fairness and recommendation performance for cold users in the context of pure CFs, i.e., the recommendation tasks are only given the user-item interaction data, in a flexible and practical manner. It can be easily applied on any recommendation model as long as it captures user preference and makes recommendations through learning user representations. To evaluate the effectiveness of our framework, we apply it over four representative recommendation models including two shallow models and two deep models. We introduce the four base models as follows:
\begin{itemize}
    \item \textbf{PMF} \cite{mnih2008probabilistic}: This algorithm adds Gaussian prior into both user and item representations for matrix factorization. 
    \item \textbf{BiasedMF} \cite{koren2009matrix}: This is a matrix factorization algorithm with user, item and global bias terms.
    \item \textbf{DMF} \cite{xue2017deep}: This is a deep model for recommendation, which uses multi-layer perceptron with non-linear activation function to encode user and item representations into dense latent factors.
    \item \textbf{MLP} \cite{cheng2016wide}: This algorithm applies deep neural network with non-linear activation functions to train a user and item latent vector matching function.
    
\end{itemize}

\subsection{\textbf{Evaluation Metrics}}
To evaluate the top-$N$ recommendation quality,
we use standard metrics Normalized Discounted Cumulative Gain ($\mathrm{NDCG}@N$) and Hit rate ($\mathrm{Hit}@N$) scores. We use sampled negative interactions for evaluation 
\cite{zhao2020revisiting, bellogin2011precision} to improve efficiency. In specific, for each user, we randomly sample 100 items which the user has never interacted with. We put those negative items together with the positive one in the validation or test set to constitute the user's candidates list. We compute the metric scores over the candidates list to evaluate the top-$N$ ranking performance. The result of all metrics in our experiments are averaged over all users.

To evaluate fairness, as we discussed above, counterfactually fair recommendation essentially requires the independence between user sensitive features and user embeddings. Therefore, we follow the settings in learning fair representations through adversary learning such as \cite{elazar2018adversarial, arduini2020adversarial, li2021towards} to train an attacker, which has totally the same structure and capacity as the discriminator. In specific, after we finish training the algorithm and getting the learned user embeddings, we train an attacker to classify the sensitive features from the user embeddings. For training and evaluating the fairness attacker, we split 80\% users into the training set and 20\% as the test set.
We report AUC score of the attacker to show if the user embedding can be classified correctly by the attacker. We say that the sensitive information is leaked into user embeddings if the attacker can successfully distinguish sensitive features from user embeddings. To meet the requirement of counterfactual fairness, an ideal result is an AUC score of about 0.5, which means that the attacker cannot distinguish the sensitive feature out of the user embedding at all.

\subsection{Experimental Results}
\label{6.5}

\begin{table*}[htbp]
  \centering
  \caption{The results of baselines and our method (TFR) under unsupervised and supervised cases on MovieLens. All the metrics are evaluated on target users. The results in the first three rows are from baselines while the last two rows are from our method. The best recommendation performance results are highlighted in bold. The best fairness performance is underlined. ``*'' means the best performance is significantly better than all the baselines based on paired $t$-test at the significance level of 0.05.}
    \begin{tabular}{lcccccccccc}
    \toprule
          & \multicolumn{5}{c}{PMF}               & \multicolumn{5}{c}{BiasedMF} \\
\cmidrule(lr){2-6}      
\cmidrule(lr){7-11}
& \multicolumn{1}{c}{NDCG@5} & \multicolumn{1}{c}{NDCG@10} & \multicolumn{1}{c}{Hit@5} & \multicolumn{1}{c}{Hit@10} & \multicolumn{1}{c}{AUC} & \multicolumn{1}{c}{NDCG@5} & \multicolumn{1}{c}{NDCG@10} & \multicolumn{1}{c}{Hit@5} & \multicolumn{1}{c}{Hit@10} & \multicolumn{1}{c}{AUC} \\
    \midrule
    Data: Source + Target & 0.3082 & 0.3568 & 0.4548 & 0.6056 & 0.6768 & 0.2886 & 0.3505 & 0.4242 & 0.6147 & 0.6526 \\
    Data: Target & 0.1067 & 0.1416 & 0.1707 & 0.2792 & 0.5663 & 0.1983 & 0.2524 & 0.3049 & 0.4739 & 0.5800 \\
    Data: Source  & 0.0796 & 0.1049 & 0.1243 & 0.2030 & 0.5000 & 0.1395 & 0.1833 & 0.2220 & 0.3596 & 0.5000 \\
    \textbf{TFR-Unsupervised} & 0.2385 & 0.2965 & 0.3554 & 0.5360 & \underline{0.5000*}   & 0.2182 & 0.2723 & 0.3322 & 0.5004 & \underline{0.5316*} \\
    \textbf{TFR-Supervised} & \textbf{0.3159*} & \textbf{0.3673*} & \textbf{0.4590} & \textbf{0.6189*} & 0.5495 & \textbf{0.3125*} & \textbf{0.3662*} & \textbf{0.4548*} & \textbf{0.6214*} & 0.5335 \\
    \midrule
          & \multicolumn{5}{c}{DMF}               & \multicolumn{5}{c}{MLP} \\
\cmidrule(lr){2-6}      
\cmidrule(lr){7-11}
& \multicolumn{1}{c}{NDCG@5} & \multicolumn{1}{c}{NDCG@10} & \multicolumn{1}{c}{Hit@5} & \multicolumn{1}{c}{Hit@10} & \multicolumn{1}{c}{AUC} & \multicolumn{1}{c}{NDCG@5} & \multicolumn{1}{c}{NDCG@10} & \multicolumn{1}{c}{Hit@5} & \multicolumn{1}{c}{Hit@10} & \multicolumn{1}{c}{AUC} \\
    \midrule
    Data: Source + Target & 0.2834 & 0.3396 & 0.4217 & 0.5965 & 0.6316 & 0.2881 & 0.3487 & 0.4275 & 0.6147 & 0.6263 \\
    Data: Target & 0.1795 & 0.2368 & 0.2833 & 0.4590 & 0.5579 & 0.2301 & 0.2759 & 0.3488 & 0.4896 & 0.6347 \\
    Data: Source  & 0.0355 & 0.0542 & 0.0630 & 0.1218 & 0.5000   & 0.0968 & 0.1407 & 0.1632 & 0.2999 & 0.5000 \\
    \textbf{TFR-Unsupervised} & 0.2207 & 0.2730 & 0.3256 & 0.4880 & \underline{0.5021*} & 0.2326 & 0.2897 & 0.3496 & 0.5269 & 0.5274 \\
    \textbf{TFR-Supervised} & \textbf{0.2863*} & \textbf{0.3473*} & \textbf{0.4288*} & \textbf{0.5989} & 0.5500  & \textbf{0.2942} & \textbf{0.3526} & \textbf{0.4358*} & \textbf{0.6164} & \underline{0.5211*} \\
    \bottomrule
    \end{tabular}%
  \label{tab:movielens_result}%
\end{table*}%

\begin{table*}[htbp]
  \centering
  \caption{The results of baselines and our method (TFR) under unsupervised and supervised cases on Last.FM. All the metrics are evaluated on target users. The results in the first three rows are from baselines while the last two rows are from our method. The best recommendation performance results are highlighted in bold. The best fairness performance is underlined. ``*'' means the best performance is significantly better than all the baselines based on paired $t$-test at the significance level of 0.05.}
    \begin{tabular}{lcccccccccc}
    \toprule
          & \multicolumn{5}{c}{PMF}               & \multicolumn{5}{c}{BiasedMF} \\
\cmidrule(lr){2-6}      
\cmidrule(lr){7-11}
& \multicolumn{1}{c}{NDCG@5} & \multicolumn{1}{c}{NDCG@10} & \multicolumn{1}{c}{Hit@5} & \multicolumn{1}{c}{Hit@10} & \multicolumn{1}{c}{AUC} & \multicolumn{1}{c}{NDCG@5} & \multicolumn{1}{c}{NDCG@10} & \multicolumn{1}{c}{Hit@5} & \multicolumn{1}{c}{Hit@10} & \multicolumn{1}{c}{AUC} \\
    \midrule
    Data: Source + Target & 0.2424 & 0.2861 & 0.3499 & 0.4843 & 0.5990 & 0.2498 & 0.2969 & 0.3610 & 0.5046 & 0.6295 \\
    Data: Target & 0.0410 & 0.0607 & 0.0645 & 0.1271 & 0.5553 & 0.0531 & 0.0727 & 0.0829 & 0.1446 & 0.6271 \\
    Data: Source  & 0.0315 & 0.0452 & 0.0497 & 0.0921 & 0.5000  & 0.0590 & 0.0851 & 0.0939 & 0.1759 & 0.5000 \\
    \textbf{TFR-Unsupervised} & 0.0651 & 0.0918 & 0.1059 & 0.1897 & \underline{0.5000*}   & 0.0847 & 0.1106 & 0.1317 & 0.2127 & \underline{0.5000*} \\
    \textbf{TFR-Supervised} & \textbf{0.2605*} &\textbf{ 0.2993*} & \textbf{0.3702*} & \textbf{0.4899} & 0.5394 & \textbf{0.2790*} & \textbf{0.3234*} & \textbf{0.3867*} & \textbf{0.5249*} & 0.5577 \\
    \midrule
          & \multicolumn{5}{c}{DMF}               & \multicolumn{5}{c}{MLP} \\
\cmidrule(lr){2-6}      
\cmidrule(lr){7-11}
& \multicolumn{1}{c}{NDCG@5} & \multicolumn{1}{c}{NDCG@10} & \multicolumn{1}{c}{Hit@5} & \multicolumn{1}{c}{Hit@10} & \multicolumn{1}{c}{AUC} & \multicolumn{1}{c}{NDCG@5} & \multicolumn{1}{c}{NDCG@10} & \multicolumn{1}{c}{Hit@5} & \multicolumn{1}{c}{Hit@10} & \multicolumn{1}{c}{AUC} \\
    \midrule
    Data: Source + Target & 0.1945 & 0.2403 & 0.2956 & 0.4374 & 0.6506 & 0.1862 & 0.2372 & 0.2799 & 0.4383 & 0.6210 \\
    Data: Target & 0.0356 & 0.0572 & 0.0562 & 0.1243 & 0.5652 & 0.0479 & 0.0674 & 0.0773 & 0.1381 & 0.5807 \\
    Data: Source  & 0.0246 & 0.0392 & 0.0378 & 0.0838 & 0.5000   & 0.0551 & 0.0791 & 0.0906 & 0.1654 & 0.5000 \\
    \textbf{TFR-Unsupervised} & 0.0339 & 0.0478 & 0.0562 & 0.1004 & \underline{0.5000*}   & 0.0702 & 0.0931 & 0.1169 & 0.1878 & \underline{0.5000*} \\
    \textbf{TFR-Supervised} & \textbf{0.2312*} & \textbf{0.2735*} & \textbf{0.3324*} & \textbf{0.4641*} & 0.5183 & \textbf{0.2313*} & \textbf{0.2816*} & \textbf{0.3490*} & \textbf{0.5055*} & 0.5272 \\
    \bottomrule
    \end{tabular}%
  \label{tab:lastfm_result}%
  \vspace{-5pt}
\end{table*}%

In this section, we show and analyze the main experimental results.
To make fair comparison and maintain the consistency of the information used by models, we train base models in three different settings.
To evaluate the effectiveness of our method under unsupervised setting, we only use the data of the source users to train the base models, and evaluate their performance on the target users. To evaluate the effectiveness of our method under supervised setting, we use the data from both source users and target users to train the base models, and evaluate their performance on the target users. We also evaluate the performance of the base models when they are trained directly on target users, and compare their performance with our model to show the power of transfer learning. We show NDCG, Hit results, and AUC scores of all the models on the MoiveLens-1M and Last.FM datasets in Table \ref{tab:movielens_result} and Table \ref{tab:lastfm_result}, respectively. We will analyze the results in the following.

\subsubsection*{\normalfont \textbf{Training Loss}} 
\begin{figure}
    \begin{subfigure}[t]{0.49\linewidth}
        \centering
        \includegraphics[scale=0.27]{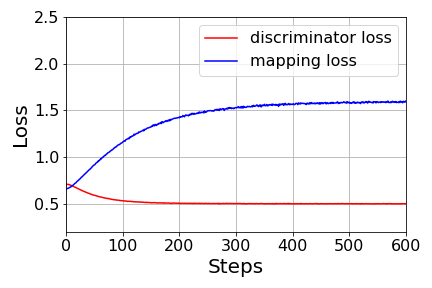}
        \caption{unsupervised training}
        \label{fig:train_loss_a}
    \end{subfigure}
    \begin{subfigure}[t]{0.49\linewidth}
        \centering
        \includegraphics[scale=0.27]{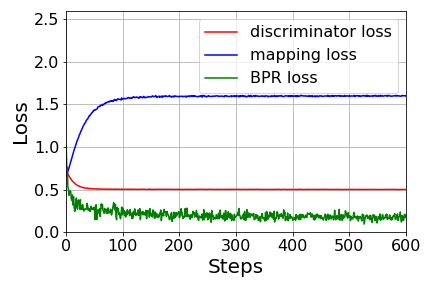}
        \caption{supervised training}
        \label{fig:train_loss_b}
    \end{subfigure}
    \caption{The adversarial training losses of using MLP model under supervised and unsupervised setting on Last.FM}
    \label{fig:train_loss}
\end{figure}

Fig.~(\ref{fig:train_loss}) shows an example of the unsupervised and supervised training losses on the Last.FM dataset. From Fig.~(\ref{fig:train_loss_a}), we see that both mapping function and discriminator converges in around 400 adversarial iterations. In Fig.~(\ref{fig:train_loss_b}), we observe the consistent trend for both adversarial losses. Additionally, the BPR loss also converges after 200 iterations. This result indicates that the target user training process can get benefits from both transfer learning and the target user interactions, which shows the effectiveness of our design.

\subsubsection*{\normalfont \textbf{Performance of Transferring Fairness}}
We first analyze if our model can achieve transferable counterfactual fairness across users. To this end, we show the AUC scores of attackers on the gender feature over all models on MoiveLens-1M and Last.FM datasets in Table \ref{tab:movielens_result} and Table \ref{tab:lastfm_result}, respectively. The best fairness performance results are underlined in the tables. It should be noted that the best fairness performance (AUC) is selected by comparing all models except for that trained only on the source data (i.e., Data: Source). The reason is that if we only train the model on the source data and then test it on the target users, the target user embeddings will just be randomly initialized vectors since they are not optimized at all, which will trivially lead to low AUC scores (i.e., 0.5). The low AUC scores do indicate that the user embeddings do not involve sensitive information since they are just random vectors, however, it is a trivial fairness achievement since the random embeddings are not informative at all to make good recommendations.

Firstly, we can see that when baseline models are trained with both source and target data (Data: Source + Target), the AUC scores are significantly higher than 0.5. It means that the sensitive information is embedded into user representations which violates the requirement of counterfactual fairness. We can also observe that when the baseline models are trained only on source (Data: Source) or target data (Data: Target), the AUC scores are much lower than when they are trained based on all data, and the AUC scores are very close to 0.5 especially when baselines are trained using only source data (Data: Source). However, it is worth noting that the low AUC scores under such case is because that the user embeddings are not fully trained, and the randomness in such embeddings leads to low AUC scores, which is shown by its poor recommendation performance. Furthermore, we can see from the tables that the AUC scores of our methods (TFR) under two settings are around 0.5, i.e., it is difficult for the attackers to distinguish the sensitive feature from the learned user embeddings. The AUC scores of unsupervised case are usually lower than supervised case, because it will be easier for the model to learn user sensitive information from target user interaction data under the supervised case, so that it is harder for the supervised case to enhance fairness. Under the supervised case, we can see that our method not only achieves low AUC scores, but also maintains desirable recommendation performance. It indicates that the learned user embeddings are informative enough to generate high-quality recommendations, and meanwhile do not contain sensitive information of the users. Therefore, the results show that our method is effective for transferring counterfactual fairness across source and target users for recommendation.

\subsubsection*{\normalfont \textbf{Performance of Unsupervised Case}}
Besides enhancing the fairness for target users, we also analyze if our model can improve recommendation quality for target users. We first compare the recommendation performance of the baseline models and our method under unsupervised case. In such case, we only use the data of source users to train a fair recommender system, and then transfer the knowledge to the target users.
It is worth noting that our method does not touch any interaction data or metadata of target users. We can see from the results that our method (TFR-Unsupervised) performs much better than the baseline models trained on source data only (Data: Source), i.e., using the same information as ours. These results show that it is effective to improve the recommendation performance of target users through transfer learning. What's more, our method performs better than most of the baseline models trained on target data only (Data: Target). For the DMF model on Last.FM, we still get comparable performance. These results show that transferring knowledge across domains even under unsupervised case is promising to achieve better accuracy than directly train a recommendation model from the sparse data of target users. Comparing the performance of our method (TFR-Unsupervised) with directly training a model on data from both types of users (Data: Source + Target), we can see our performance is not as good as the baseline models. This is reasonable since the baseline models use much more information than our method in this case, and we will show in the following that our model will perform better than baselines when we use the same information.

\subsubsection*{\normalfont \textbf{Performance of Supervised Case}}
We also compare the recommendation quality of the baseline models and our method under supervised case. When there is interactive data available for target users, our method provides the flexibility to add such information to achieve better recommendation accuracy. We can see from the results that our method (TFR-Supervised) outperforms all baseline models on all evaluation metrics no matter what kind of data baseline models use for training, which verifies the effectiveness of our model for transferring recommendation performance across users.

\section{Conclusion}
\label{sec:conclusions}
In this paper, we study the problem of transferring recommendation accuracy and counterfactual fairness across source and target users in recommender systems. We focus on the case where there are new users coming with limited or even no interaction to train the model. We treat the warm-start users with sufficient interaction data as the source users, while the cold-start users with few or no interaction data are treated as the target users. Our goal is to transfer the knowledge of a fair model trained on the source users to the target users to improve the recommendation performance as well as keeping the fairness property for the target users. 
To achieve this goal, we introduce a framework to transfer the user preferences and counterfactual fairness in recommendation by taking advantage of adversary learning. 
Experiments on two real-world datasets with representative shallow and deep models verified the effectiveness of our method in terms of both improving the recommendation performance and improving the recommendation fairness for the target users. 


\bibliographystyle{ACM-Reference-Format}
\bibliography{sample-base}

\clearpage
\newpage

\appendix
\section*{Appendix}
\section{Algorithm}
\label{ap:alg}
\begin{algorithm}[h]
\caption{Learning Algorithm for TFR}
\label{alg:train}
\SetKwInOut{Input}{Input}
\SetKwInOut{Init}{Initialize}
\SetKw{PhaseA}{Step 1:}
\SetKw{PhaseB}{Step 2:}
\SetKwRepeat{Repeat}{repeat}{until}
\SetAlgoLined
\Input{Source data $\mathcal{D}_S=\{u_S, v, Y_S\}$; Target data $\mathcal{D}_T=\{u_T, v, Y_T\}$; 
Fairness filter $f_A$; Domain mapping function $M$; Fairness discriminator $D_A$; Domain discriminator $D_d$; Recommender system model parameters $\Theta_s, \Theta_t$}

\BlankLine
\Init{$f_A$, $M$, $D_A$, $D_d$, $\Theta_s$, $\Theta_t$}
\BlankLine

\PhaseA{\textit{pre-train fair model on source user}}

\Repeat {stopping criterion is met}{
    \For{$u, v\in \mathcal{D}_S$}{
        $\mathcal{L}_1=\mathcal{L}_{\text {Rec}}(u,v, Y_S)-\lambda_A  \mathcal{L}_A \left(f_A (\textbf{r}_u), A\right)$
        $\Theta_s, f_A, D_A \leftarrow \min \mathcal{L}_1$
    }
}
\BlankLine
\PhaseB{\textit{train mapping function on target user}}

\uIf{unsupervised case}{
    \Repeat{stopping criterion is met}{
        \For{$u\in \mathcal{D}_S \cup \mathcal{D}_T$}{
            $D_d \leftarrow \max \log \left( D_d\left(f_A(\mathbf{r}_u^S)\right)\right)+\log \left(1-D_d\left(M\left(\mathbf{r}_u^T\right)\right)\right)$\;
            $M \leftarrow \min \log \left(1-D_d\left(M\left(\mathbf{r}_u^T\right)\right)\right)$
        }
    }
}

\uIf{supervised case}{
    \Repeat{stopping criterion is met}{
        \For{$u, v\in \mathcal{D}_S \cup \mathcal{D}_T$}{
            $\Theta_t, M \leftarrow \min_{u \in \mathcal{D}_T} \mathcal{L}_{Rec}(u, v, Y_T) + \lambda_D \log \left(1-D_d\left(M\left(\mathbf{r}_u^T\right)\right)\right)$\;
            
            $D_d \leftarrow \max \log \left( D_d\left(f_A(\mathbf{r}_u^S)\right)\right)+\log \left(1-D_d\left(M\left(\mathbf{r}_u^T\right)\right)\right)$\;
            
        }
    }
}
\label{alg}
\end{algorithm}

\section{Proof of Theorem 5.1}
\label{ap:proof}
\begin{proof}
The proof contains two parts, and are  naturally applications of the Proposition 2 of Generative Adverserisal Networks (GAN) \cite{goodfellow2014generative}, which proves that if (1) the generator and discriminator have enough capacity, (2) the discriminator is allowed to reach the optimum during the training process, and (3) the generator is updated with the discriminator fixed so as to improve the criterion, then the distribution of the fake data generated from generator will converge to the distribution of the real data. In our paper, for simplicity, we consider the case of a single binary sensitive feature, i.e., we have one sensitive feature $A$ which can be 0 or 1. It can be naturally generalized to the multi-class and multi-feature settings \cite{cho2020fair}.

Firstly, by simply replacing the task of distinguishing real and fake data with classifying binary sensitive attribute, we have that after the first step of Algorithm~\ref{alg}, the distributions of user embeddings with sensitive feature $A=0$ and $A=1$ will be the same once the algorithm converged, i.e., for any $u \in \mathcal{D}_S$, we have $P(f_A(\textbf{r}_u^S)|A=0)=P(f_A(\textbf{r}_u^S)|A=1)$, which gives $f_A(\textbf{r}_u^S) \perp A$. Secondly, after the second step of Algorithm~\ref{alg}, we have $P(f_A(\textbf{r}_u^S))=P(\textbf{r}_u^T)$ directly according to Proposition 2 of \cite{goodfellow2014generative}. 
Therefore, after combining the conclusions of the two steps, we naturally obtain that for any user $u \in \mathcal{D}_T$, $\textbf{r}_u \perp \textbf{A}_{u}$ as $\lambda\rightarrow \infty$. 
\end{proof}

\section{Data Split}
\label{ap:data}
To create cold-start users, following previous work such as \cite{chae2019rating}, we randomly sample 20\% users from the dataset, and randomly keep at most 5 interactions for each of the users to construct the target user set. We treat those users with no more than 5 interactions as the target users while keeping the rest as the source users. We remove those items that appear only in the target user set to ensure no cold item exists in the target user set.

For training and evaluating the model, we do leave-one-out to split the whole dataset into train, validation and test sets, which is a commonly used experimental setting in the literature~\cite{chen2021neural, shi2020neural, chen2021neural}. In specific, for each user, we randomly hold out two interacted items and put them into validation and test set, respectively. We do the same for both source and target users, which means that both source and target users have their independent train, validation and test sets. This is to prepare for testing the performance of baselines under different settings, which will be discussed in detail in Section \ref{6.5}, and our supervised and unsupervised experimental settings. 

\section{Experimental Settings}
\label{ap:exp}
\subsection{\normalfont \textbf{Settings for the First Step}}
We apply the Bayesian Personalized Ranking (BPR)~\cite{rendle2012bpr} loss as the recommendation loss $\mathcal{L}_{Rec}$ in Eq.\eqref{eq} for all the baseline models. For each user-item pair in the training set, we randomly sample one item that the user has never interacted with as the negative sample in one training epoch. The learning rate is 0.001.  $\ell_2$-regularization coefficient is chosen from $\{10^{-4}, 10^{-5}, 10^{-6}\}$. Early stopping is applied and the best models are selected based on the performance on the validation set. We apply Adam~\cite{kingma2014adam} as the optimization algorithm to update the model parameters. The fairness adversarial coefficient $\lambda_A$ in Eq.\eqref{eq} is selected from $\{1, 5, 10\}$ for both datasets. The filter modules are two-layer neural networks with LeakyReLU as the non-linear activation function and batch normalization layer is applied. The fairness classifiers (discriminators and attackers) are multi-layer perceptrons with the number of layers set to 6, LeakyReLU as the activation function, and the dropout rate is set to 0.3. All the model parameters are randomly initialized from a 0 mean, 0.01 standard deviation normal distribution.

\subsection{\normalfont \textbf{Settings for the Second Step}} To generate high-quality recommendations for target users, we learn embeddings for target users through adversary learning, while keeping all the other model parameters the same as those learned from the source users. For example, for the PMF model, we directly copy the item embeddings learned from source users to target users and freeze those item embeddings during training. The embeddings of source users are also frozen, and we only update the embeddings of target users during the learning process.

For the unsupervised setting, we have four-layer MLP as the domain mapping function $M$ for base models. The activation function is LeakyReLU. We apply batch normalization between each two layers except for the output layer. The discriminator is a six-layer MLP with LeakyReLU as the activation function. The dropout rate for domain discriminator $D_d$ is selected between 0.3 and 0.5. To stabilize adversarial training, we apply stochastic gradient decent (SGD) optimization algorithm for the discriminator while apply Adam as the optimizer for the mapping function $M$. This is to slow down the convergence of discriminator at the early training stage and help the generator to learn useful knowledge from discriminator. It is a commonly used strategy to balance adversarial learning process. We also apply spectral normalization~\cite{miyato2018spectral} to the discriminator to prevent the training process from failure mode, i.e. the discriminator is too strong to suppress the generator or vice versa. The learning rate of $M$ is 0.001 and the learning rate for discriminator $D_d$ is adjusted between 0.0001 and 0.001. For the supervised training, we keep most of the settings the same as unsupervised training. The empirical adversarial coefficient $\lambda_D$ between recommendation loss $\mathcal{L}_{Rec}$ and the adversarial loss $\mathcal{L}_{D}$ is between 1 and 10. 

\subsection{\normalfont \textbf{Adversarial Training}}
For adversarial training of the first step, we train the discriminator and filter function alternatively. We follow the operations in~\cite{li2021towards} by firstly feeding input data into the model to compute $\mathcal{L}_{Rec}$ and $\mathcal{L}_A$. We fix the parameters of the discriminator and optimize the recommendation model as well as the filter function by minimizing the loss $\mathcal{L}_1$. After that, we fix the parameters of the recommendation model and the filter function to update discriminator through minimizing $\mathcal{L}_A$ for $t$ steps. We use $t=10$ in our experiments.

For domain transfer training in the second step, we have the similar process as what we did in the first step. We first train the discriminator for 1 step with all the user embeddings from source and target users. Then we fix discriminator and optimize the mapping function for 1 step. We do this process repeatedly until the convergence condition is met. For the supervised setting, we train recommendation model together with adversary training similarly as the first step.

\end{document}